\documentstyle[preprint,aps]{revtex}
\begin{document}
\draft
\title{Characteristics of highly excited diatomic 
	rovibrational spectra and slow atomic 
	collisions}
\author{Bo Gao}
\address{Department of Physics and Astronomy,
   University of Toledo,
   Toledo, Ohio 43606}
\date{\today}
\maketitle
\begin{abstract}

We show that the highly excited rovibrational spectra
of a diatomic molecule and the closely related 
slow atomic collision processes contain more 
systematics and require less parameters to 
characterize than the Rydberg spectrum of an atom.
In the case of a single channel, for example, 
we show that a {\em single\/} short-range parameter gives
a complete description of slow collisions
for practically all angular momenta, and covers an energy
range of hundreds of millikelvins. The {\em same\/} 
parameter also describes the highly excited rovibrational 
spectra in the threshold region, including states of
different angular momenta.
Sample applications and predictions of the theory 
are discussed, including comparisons with 
experiment.

\end{abstract}
\pacs{PACS number(s): 34.10.+x,33.20.Vq,32.80.Pj}

\narrowtext

This paper examines the properties of a two-atom
system in a small energy region around a
dissociation threshold, with the region above
corresponding to slow atomic collisions,
and the region below containing the
highly excited rovibrational states of the
corresponding diatomic molecule.
We discuss a type of systematics that relate states
of different relative angular momenta, 
thus greatly simplifying our understanding of
such systems. Comparison of the theory with 
existing experimental data \cite{tsa97} also
provides the first verification of the breakdown
of the large-quantum-number formulation 
of the correspondence principle \cite{tro98,boi98,gao99b}.

With recent developments such as 
the analytic solutions of
the Schr\"{o}dinger equation for $1/r^6$ and $1/r^3$
potentials \cite{gao98a,gao99a,gao99b}, 
our understanding of two-atom systems is already
conceptually comparable to the quantum defect
theory (QDT) of atomic Rydberg spectra \cite{qdt}. 
Namely, the slow atomic scattering
and the rovibrational spectrum of a particular angular 
momentum can be understood in terms of the long range 
solutions and a parameter that is a slowly varying 
function of energy in the threshold region \cite{gao98a,gao98b,bur98a,gao99a,gao99b}. 
In this work, we show that a two-atom system 
possesses an even greater degree of 
systematics in the following sense. Unlike the QDT for
atomic Rydberg spectra in which different angular
momentum states have different quantum defects with
no general relationships among them, a single parameter
is sufficient to characterize slow atomic collisions
for practically all angular momenta. The same parameter
also characterizes the rovibrational spectra in
the threshold region, including states of different 
angular momenta. In other words, in addition to the relationship 
between the bound spectrum and scattering, as expected from
traditional QDT formulations, a two-atom system has also
the unique property that scattering 
of different angular momenta are related, and 
so are the bound spectra of different angular momenta.
While a similar idea can be implemented
in a numerical fashion \cite{tsa97}, it becomes a much 
more powerful tool in the present formulation.

The origin of this systematics is not difficult to
understand and is due to a combination of the following
three properties of a typical molecular system. 
(a) Atoms are strongly repulsive at
short distances. 
(b) Atoms are heavy compared to electrons. 
(c) The atom-atom interaction at large distances behaves
as $1/r^n$ with $n>2$.

The combination of the first two properties 
gives rise to the well-known separation of tightly bound
rovibrational states into the product of two parts 
$\psi_v\psi_r$, in which the vibrational wave function 
$\psi_v$ is, to the lowest order, independent of rotational
quantum numbers. 
For rovibrational states that are highly excited, 
the same properties ensure that the radial wave function, 
up to a normalization constant, remain nearly independent 
of the angular momentum until a distance
is reached where the rotational energy term becomes
comparable to the electronic and other energy terms.
This characteristic is uniquely molecular
and does not apply to the electronic wave functions of an atom, 
for which the rotational term dominates the behavior at 
sufficiently small $r$.

The implication of the third property can be understood from
qualitative behaviors of the effective potential
$-C_n/r^{n}+\hbar^2 l(l+1)/(2\mu r^2)$, in which the
first term represents the long range interaction and
the second term represents the centrifugal barrier.
For $n>2$ and $l\neq 0$, this potential has a maximum 
and crosses zero at
$
r_x = \left[l(l+1)\right]^{-1/(n-2)}\beta_n ,
$
where $\beta_n$ is a length scale associated with
the $-C_n/r^n$ interaction as defined by
\begin{equation}
\beta_n \equiv (2\mu C_n/\hbar^2)^{1/(n-2)} \;.
\label{eq:betan}
\end{equation}
From a physical point of view, $r_x$ specifies the location at 
which the rotational potential is equal to the long range 
interaction. For an $l$ that is not too large, 
$r_x$ is of the order of $\beta_n$.   
Since $n$ is greater than two, for distances that
are smaller than $r_x$, the long range interaction
quickly dominates, and correspondingly, the importance 
of the angular momentum quickly diminishes.
Mathematically speaking, this means that 
for a potential that goes to zero faster than $1/r^2$ at
large distances, a pair of linearly independent solutions 
exist, which, at relatively small distances, are not only 
independent of energy, but also independent of angular momentum.

The combination of these characteristics
leads to the following important conclusion.
With proper choice of long range solutions,
a quantum defect theory for molecular
rovibrational states and slow atomic collisions 
can be formulated in which the short-range parameters 
are not only nearly independent of energy, but also nearly 
independent of the relative angular momentum.
This conclusion is very general and powerful.
Its formulation and application are illustrated
here for a single channel problem with a van der Waals
asymptotic interaction. The formulation for attractive 
$1/r^3$ interaction and multichannel generalizations 
will be presented elsewhere \cite{gao00a}.
 
For a single channel with an asymptotic 
van der Waals interaction ($-C_6/r^6$),
the proper pair of long range solutions 
are ones with the behavior
\begin{eqnarray}
f^c_{\epsilon l} &\stackrel{r\ll\beta_6}{\longrightarrow}& 
	(2/\pi)^{1/2}
	(r/\beta_6)r^{1/2}\cos\left(
	\frac{1}{2}(r/\beta_6)^{-2}-\frac{\pi}{4} \right) \;, 
\label{eq:fcasy0}\\
g^c_{\epsilon l} &\stackrel{r\ll\beta_6}{\longrightarrow}& 
	-(2/\pi)^{1/2}
	(r/\beta_6)r^{1/2}\sin\left(
	\frac{1}{2}(r/\beta_6)^{-2}
	-\frac{\pi}{4} \right) \;,
\label{eq:gcasy0}
\end{eqnarray}
for both positive and negative energies.
This pair, which has not only energy-independent, but
also angular-momentum-independent behavior at small distances,
is related to the $f^0$ and $g^0$ pair 
defined in \cite{gao98a} by a 
linear transformation. 
The fact that such a pair exist
is a result of the exponent $n$ that characterizes the long
range interaction, $-C_n/r^n$, being greater than 2.

With this choice of long range solutions,
the single channel quantum defect theory is still formally
the same as discussed previously \cite{gao98b,gao99a,gao99b}.
In particular, the bound spectrum of any potential with the
behavior of $V(r)\rightarrow -C_6/r^6$ at large distances is 
still given rigorously by the crossing points 
between a $\chi$ function and a short-range 
K matrix \cite{gao98b,gao99b}:  
\begin{equation}
\chi^c_l(\epsilon_s) = K^c(\epsilon,l) \;.
\label{eq:EnergyLevels}
\end{equation}
Here $\epsilon_s$ is a scaled bound state energy defined by
\begin{equation}
\epsilon_s = \frac{1}{16}\frac{\epsilon}{(\hbar^2/2\mu)(1/\beta_6)^2} \;.
\label{eq:es6}
\end{equation}
The $\chi^c$ function, corresponding to the choice
of long range solutions as specified by
Eqs.~(\ref{eq:fcasy0}) and (\ref{eq:gcasy0}), 
is given by
\begin{equation}
\chi^c_l(\epsilon_s) = 
	\frac{(Y_{\epsilon l}/X_{\epsilon l})+
	\tan(\pi\nu/2)(1+M_{\epsilon l})/(1-M_{\epsilon l})}
	{1-(Y_{\epsilon l}/X_{\epsilon l})
	\tan(\pi\nu/2)(1+M_{\epsilon l})/(1-M_{\epsilon l})} \;,
\label{eq:chi3}
\end{equation}
in which $M_{\epsilon l}=G_{\epsilon l}(-\nu)/G_{\epsilon l}(\nu)$,
with $\nu$ and $G_{\epsilon l}$ being defined in \cite{gao98a}. 
Plots of $\chi^c_l$ functions
for different $l$ are shown in Figure~\ref{fig:chic6}.
$K^c$ is the short range K matrix that results from
the matching of the short range solution and the long
range solution. It is given specifically by
\begin{equation}
K^c(\epsilon, l) = \left(\frac{f^c_{\epsilon l}}{g^c_{\epsilon l}}\right)
	\frac{({f^c_{\epsilon l}}'/f^c_{\epsilon l})
	- ({u_{\epsilon l}}'/u_{\epsilon l})}
	{({g^c_{\epsilon l}}'/g^c_{\epsilon l})
	- ({u_{\epsilon l}}'/u_{\epsilon l})} \;,
\label{eq:Kc}
\end{equation}
evaluated at any radius beyond $r_0$ at which the 
potential has become well represented by $-C_n/r^n$.

Above the threshold, the scattering $K$ matrix is 
given by a similar expression as derived 
previously \cite{gao98b,gao99a,gao99b}:
\begin{equation}
K_l \equiv \tan\delta_l = 
	(K^c Z^c_{gg}-Z^c_{fg})(Z^c_{ff}-K^c Z^c_{gf})^{-1} \;,
\label{eq:qdtpe}
\end{equation}
with the $Z^c$ matrix being related to the $Z$ matrix
defined in \cite{gao98a} by a linear transformation.

The key difference between the present formulation and the
previous one is the following. With the choice of
long range solution pair as specified by 
Eqs.~(\ref{eq:fcasy0}) and (\ref{eq:gcasy0}), and the property of
the wave function discussed earlier, the parameter $K^c$ is, under 
the condition of $\beta_n \gg r_0$ \cite{gao98b,gao99a,gao99b},
not only independent of energy, but also independent of
angular momentum $l$ in the threshold region [see Eq.~(\ref{eq:Kc})].
In other words, a single parameter, $K^c$, can describe scattering
and bound states of different angular momenta.

The determination of the parameter $K^c$ is,
in principle, straightforward theoretically. 
It can be done, for example, by solving the radial
equation at a small energy (include zero energy) and 
matching it to the proper long range solution. 
However, since our
present knowledge about both the short range 
interaction and the $C_n$ coefficients is still not
sufficiently accurate for many systems, we focus
here on the direct experimental determination of both 
the $K^c$ parameter and the $C_n$ coefficient. 
In particular, we show that 
if the $C_n$ coefficient that characterizes
the dominant long range interaction is known  
accurately, the parameter $K^c$ can be obtained 
from the measurement of a single binding energy 
(similar to the determination of $K^0_l$ discussed
previously \cite{gao98b}). 
If binding energies of more than one state are known,
in addition to the prediction of the parameter $K^c$, 
an accurate determination of $C_n$ can also be made,
especially when the two states are closely spaced
in energy.

In Table~\ref{tb:Rb851}, the first column presents
the experimental results \cite{tsa97} for states 
of $^{85}$Rb$_2$ characterized by
quantum numbers $F_1=3$, $F_2=3$, $F=6$, $l=2$, and
$T=8$ \cite{qns,gao96}. The results in the column labeled by Theory
II represents a simple calculation making use of 
only a single experimental energy.
In this calculation, the energy of the level labeled
by $v_{max}-v = 1$ in Table~\ref{tb:Rb851} 
is first scaled according to Eqs.~(\ref{eq:es6}) 
and (\ref{eq:betan}) with $\mu =77392.368$ a.u. 
and a value of $C_6=4426$ a.u. from the 
theoretical calculation of Marinescu {\it et al.\/}
\cite{mar94}. The parameter $K^c$ is determined by 
evaluating the $\chi^c_{l=2}$ function at this scaled 
energy \cite{gao98b}, which yields a value of
$K^c=0.2841$. 
The crossing points of this constant with 
$\chi^c_{l=2}(\epsilon_s)$ give the predictions
of the other energy levels listed in this column 
(see Fig.~\ref{fig:chic6}).
Note that this simple calculation already yields excellent
agreement with experiment and compares very favorably
with a much more complex numerical calculation \cite{tsa97}
listed under Theory I in Table~\ref{tb:Rb851}.
This excellent agreement shows that the parameter 
$K^c$ is indeed, to a very good approximation, a constant 
in the energy range covered by the experiment ($\sim 10$ GHz). 
This is further confirmed by the fact that enforcing
$K^c$ to be constant leads to an even better agreement, 
a procedure that demonstrates
a new method for the experimental determination
of the leading $C_n$ coefficient from a minimum of
only two binding energy measurements. 

Keeping in mind that the $C_6$ value used 
in the first calculation is not necessarily the true 
$C_6$, it is allowed to vary in the following calculation.
With more than one experimental binding energy
available, we vary $C_6$ in such a way that the
two $\chi^c$'s evaluated at the two different scaled 
energies (they may also correspond to different $l$.)
are the same, in other words, we force the parameter $K^c$ to 
be a constant. The refined $C_6$ is then given by
the root of the following equation
\begin{equation}
\chi^c_l({\epsilon_{s1}})-\chi^c_{l'}({\epsilon_{s2}})=0 \;,
\label{eq:findC6} 
\end{equation}
where $\epsilon_{s1}$ and $\epsilon_{s2}$ are two
experimental energies scaled according to
Eqs.~(\ref{eq:es6}) and (\ref{eq:betan}) 
for van der Waals interactions.
The reduced masses are known with great precision
from atomic masses so that the only unknown on
the left hand side is the $C_6$ coefficient.
This procedure, when applied to the two $d$ states labeled
in Table~\ref{tb:Rb851} by
$v_{max}-v = 1$ and 2, respectively, 
leads to a revised 
$C_6=4533$ a.u., which is in good agreement with 
$C_6=4550\pm 100$ a.u., determined by 
Boesten {\it et al.\/} \cite{boe96}.
This $C_6$ value refines the energy scaling and leads
to a revised value of $K^c = 0.3106$.
The column in Table~\ref{tb:Rb851} labeled
as Theory III gives the result of energy levels predicted
using this set of revised parameters, and an even better
agreement with experiment is achieved.
This procedure demonstrates a new method for the
determination of the $C_n$ coefficient from two or more
values of experimental binding energies.
It requires no knowledge of the short range interactions,
nor does it require numerical solutions of the
Schr\"{o}dinger equation. The closer in energy those 
two states are, the better this method works. 

As stated earlier, the parameter $K^c$ describes 
not only the rovibrational states of a particular 
relative angular momentum, but also the other angular momentum
states in the threshold region (see Fig.~\ref{fig:chic6}).
Table~\ref{tb:Rb852} gives the predictions of the bound
spectra of $^{85}$Rb$_2$ characterized by quantum numbers  
$F_1=3, F_2=3, F=6, l, T=F+l$, with
$l=0,2,4,6$, respectively \cite{qns}. They are predicted
using parameters $C_6=4533$ a.u. and $K^c = 0.3106$.

The fact that $K^c$ is nearly a constant in
the threshold region has also the 
following important implications. 
(a) It provides an experimental
verification of the breakdown of the large-quantum-number
formulation of the correspondence
principle, as a semiclassical theory would have predicted
different results for, e.g., the spacing between scaled
energies \cite{gao99b}. 
(b) Since 1 GHz corresponds to 47.99 mK, the fact that $K^c$
is independent of energy over a range of 10 GHz 
below the threshold implies
that the same constant $K^c$ can describe collisions
over hundreds of millikelvins above the threshold
(for all $l$'s that are not too large). (Similar conclusion
can also be derived for Li \cite{gao98b,gao00a}).
All the energy and angular momentum dependences in this 
energy range, including shape
resonances, are determined by the dominant long range 
interaction, in combination with the centrifugal barrier.
And they are described analytically by the
solutions for the long range potentials \cite{gao98a,gao99a,gao99b}.
This result, along with the fact that 
for small energies, the scattering for large angular momentum
is nearly independent of the value of $K^c$ and
is determined entirely by the long range interaction,
means also that the same parameter describes 
slow collisions for practically all angular momenta.
As one implication, a single short-range parameter 
(a few parameters in the case of multichannels \cite{gao00a}) 
describes not only the ultracold region of energies, but also
the region of evaporative cooling and the region of
laser cooling. In contrast, in the traditional effective 
range expansion,
a single parameter, the scattering length, gives little 
more than the cross section at zero energy. 
Figure~\ref{fig:Rb85pxs} shows the partial scattering cross sections
for $s$, $d$, and $g$ waves predicted by the same short
range parameter $K^c = 0.3106$ for two spin-polarized
$^{85}$Rb atoms in state $F_1=3$, $M_{F1}=3$,
$F_2=3$, and $M_{F2}=3$.
Only a small energy range is shown here to make for
easy identification of the narrow $g$ wave shape resonance,
observed previously by Boesten {\it et al.\/} \cite{boe96}.
More data on the scattering cross sections, including
those for larger angular momenta and over a greater
energy range, and the summation formulae for summing over 
all angular momentum
states will be presented elsewhere \cite{gao00a}.
(c) Since the hyperfine splitting is typically of the
order of 1 GHz, it means that the frame-transformation 
method for atom-atom scattering \cite{gao96}, 
when properly formulated with $K^c$ 
(or the $K^0_l$ \cite{gao98b,gao99a}) being the short range
parameter \cite{gao00a}, should work well in a multichannel 
formulation that includes hyperfine structures \cite{gao96,bur98a}. 

In conclusion, a theory of slow atomic collisions 
and molecular rovibrational states has been formulated
that takes full advantage of
the intrinsic characteristics of a typical molecular system.
This formulation establishes a simple relationship 
between states of different relative angular momenta
and greatly reduces the number
of parameters required to describe slow atomic collisions
and highly excited molecular rovibrational spectra.
Multichannel generalizations do not change the key concepts
in any substantial way, and the theory can also be implemented
in exact numerical calculations. As far as parameterization
is concerned, the theory can be further extended to an even 
greater range of energies by, e.g., taking into account the energy
and angular momentum dependences induced by the correction
terms to the dominant long range interaction \cite{corr}.
Finally, the good agreement between the theory and experiment
provides the first experimental confirmation of the
breakdown of the large-quantum-number formulation of the
correspondence principle \cite{tro98,boi98,gao99b}.

%

I thank Anthony F. Starace and Tom Kvale for careful 
reading of the manuscript and for helpful discussions.
This work is supported by National Science Foundation
under Grant No. PHY-9970791.

%
%

%

%
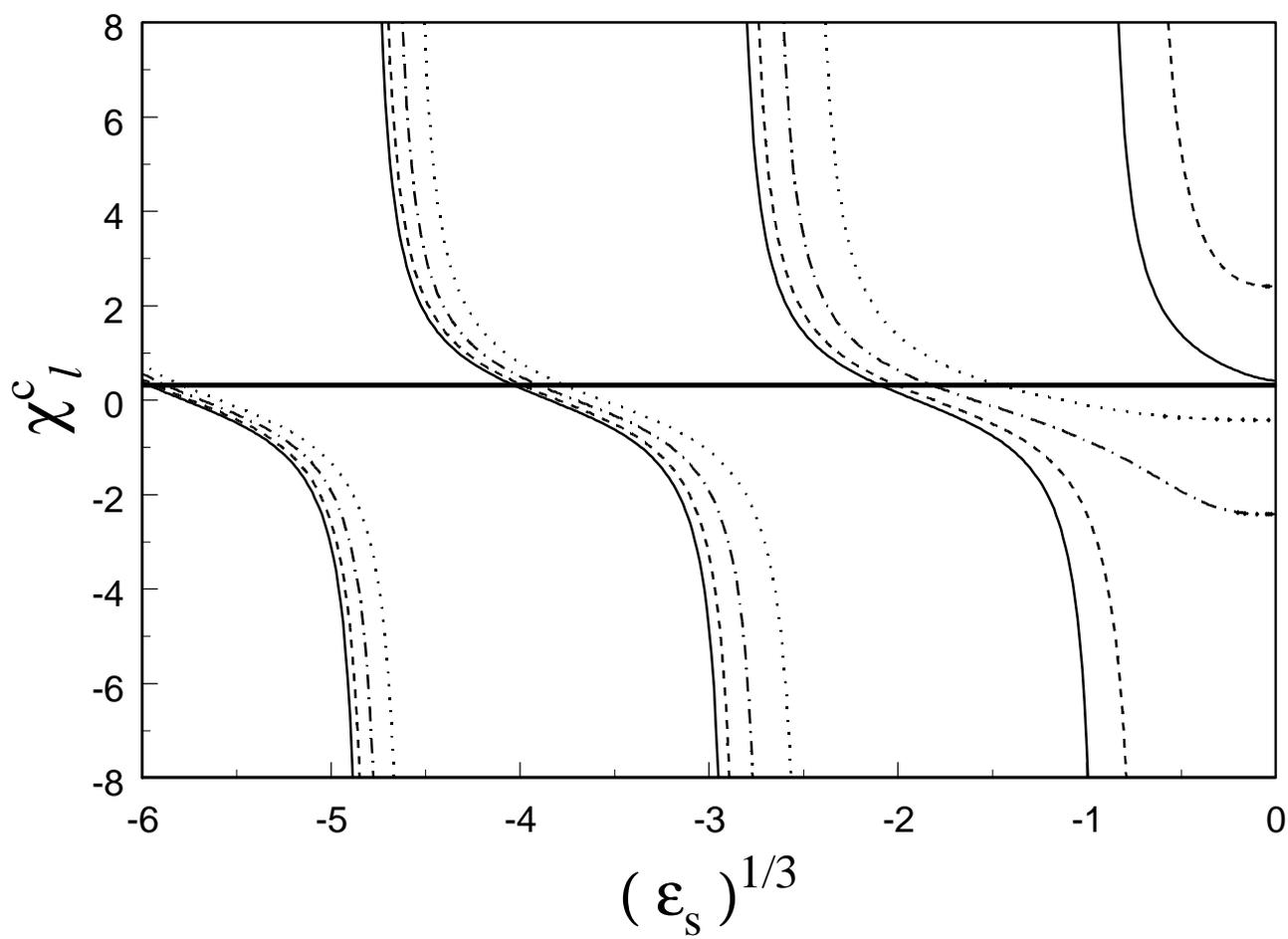
\begin{figure}
\caption{The $\chi^c_l$ functions for an attractive $1/r^6$ 
interaction plotted vs $(\epsilon_s)^{1/3}$. 
Solid line: $l=0$; Dash line: $l=1$; Dash-dotted line: $l=2$
dotted line: $l=3$. 
The bound spectra of any potential
with $V(r)\rightarrow -C_6/r^6$ at large distances is given by the 
crossing points of this {\em same\/} set of functions with a set of
short range parameters $K^c(\epsilon,l)$. 
For systems that satisfy $\beta_6\gg r_0$, 
$K^c(\epsilon,l)$ is approximately an {\em l-independent\/}
constant in the threshold region, and the bound
spectra of different angular momenta are given by the crossing
points of $\chi^c_l$ functions with a single horizontal
line representing $K^c=const.$
The solid horizontal line in this figure corresponds to
$K^c = 0.3106$. Its crossing points with $\chi^c_l$, along
with the energy scaling factor determined by $C_6=4533$ a.u.
and $\mu =77392.368$ a.u., give the binding energies
of $^{85}$Rb$_2$ presented in the last columns of 
Table~\ref{tb:Rb851} and Table~\ref{tb:Rb852}.}
\label{fig:chic6}
\end{figure}
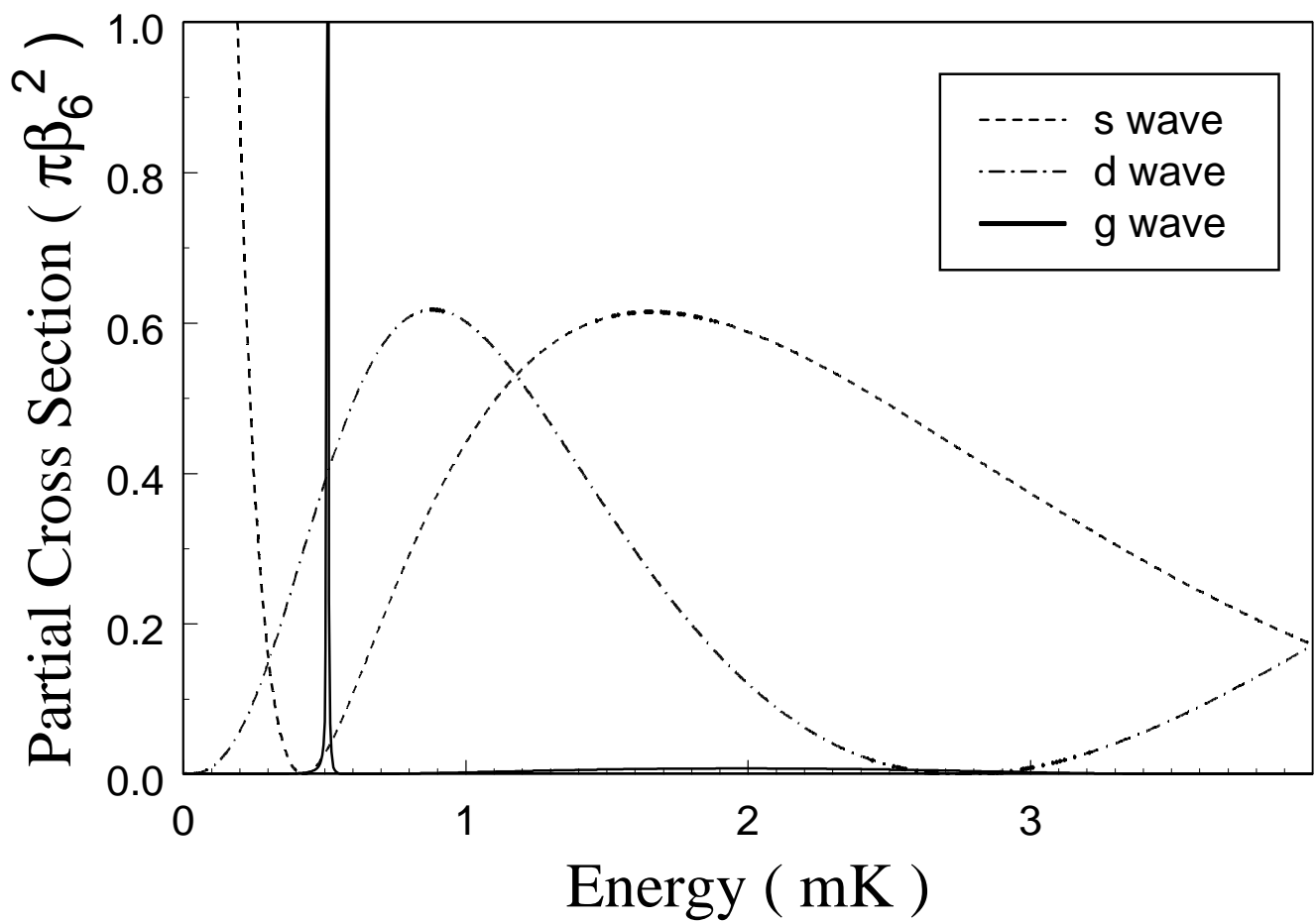
\begin{figure}
\caption{Partial scattering cross sections for a pair of
spin-polarized $^{85}$Rb atoms in state $F_1=3$, $M_{F1}=3$,
$F_2=3$, and $M_{F2}=3$. $\beta_6 = 162.7$ a.u. is determined with
$\mu =77392.368$ a.u. and $C_6 = 4533$. } 
\label{fig:Rb85pxs}
\end{figure}

\begin{table}
\caption{Comparison of energies, in GHz,  
	of the last four bound states of $^{85}$Rb$_2$ 
	characterized by quantum numbers 
	$F_1=3, F_2=3, F=6, l=2, T=8$.}
\begin{tabular}{ccccc}
	  $v_{max}-v$ 
	& Experiment\tablenote{From \cite{tsa97}.}
	& Theory I  \tablenote{From \cite{tsa97}.}
	& Theory II \tablenote{$C_6=4426$ a.u. and $K^c = 0.2841$.} 
	& Theory III\tablenote{$C_6=4533$ a.u. and $K^c = 0.3106$.}\\
\tableline
0 & $-0.16\pm 0.03$  & -0.15  & -0.1513 & -0.1539 \\
1 & $-1.52\pm 0.03$  & -1.50  & -1.520  & -1.520 \\
2 & $-5.20\pm 0.03$  & -5.16  & -5.222  & -5.200 \\
3 & $-12.22\pm 0.06$ & -12.21 & -12.37  & -12.30
\end{tabular}
\label{tb:Rb851}
\end{table}
\begin{table}
\caption{Bound state energies (in GHz)  
	of $^{85}$Rb$_2$ with quantum numbers  
	$F_1=3, F_2=3, F=6, l, T=F+l$ with
	$l=0,2,4,6$, respectively. They are predicted
	using $C_6=4533$ a.u. and $K^c = 0.3106$.}
\begin{tabular}{cccc}
	  $l=0$ 
	& $l=2$
	& $l=4$
	& $l=6$ \\
\tableline
-0.2341 & -0.1539 &        & \\
-1.678  & -1.520  & -1.161 & -0.6244 \\
-5.434  & -5.200  & -4.659 & -3.826  \\
-12.61  & -12.30  & -11.58 & -10.46  \\
-24.30  & -23.91  & -23.01 & -21.61
\end{tabular}
\label{tb:Rb852}
\end{table}
\end{document}